%% file: main.tex
\documentclass[conference]{IEEEtran}
\IEEEoverridecommandlockouts

\usepackage{graphicx}
\usepackage{subfigure}
\usepackage{amsmath,amsthm,amsfonts,amssymb}
\usepackage{cite}
\usepackage{bm}
\usepackage{bbm}
\usepackage{url}
\usepackage{array}
\usepackage{color,soul}
\usepackage{multirow}
\usepackage{booktabs}
\usepackage[table,xcdraw]{xcolor}
\usepackage{enumitem}
\usepackage{verbatim}
\usepackage{amsmath}
\usepackage{mathrsfs} 
\usepackage{xcolor}
\usepackage{setspace}

\theoremstyle{plain}
\newtheorem{thm}{Theorem}

\newtheorem{prop}[thm]{Proposition}

\newtheorem{defi}[thm]{Definition}
\newtheorem{rem}{Remark}

\newenvironment{NewProof}{{\noindent\it Proof.}}{\hfill $\blacksquare$\par}

\usepackage[english]{babel}
\usepackage{algorithm}
\usepackage[noend]{algpseudocode}
\begin{document}

\title{Resolving the Discontinuity of Continuous-Time AFDM Waveforms}
\author{Yewen Cao and Yulin Shao
\thanks{Y. Cao is with the State Key Laboratory of IoT for Smart City, University of Macau. She is also with the Department of Electrical and Computer Engineering, The University of Hong Kong  (yc47409@um.edu.mo).}
\thanks{Y. Shao is with the Department of Electrical and Computer Engineering, The University of Hong Kong (ylshao@hku.hk).}
}

\maketitle

\begin{abstract}
Continuous-time affine frequency division multiplexing (AFDM) waveforms, constructed via frequency wrapping and phase correction, are known to be sample-wise equivalent to the widely adopted discrete AFDM framework. In this paper, we uncover a fundamental and previously overlooked flaw in this construction: its complex envelope is inherently discontinuous for generic chirp parameters. We show that these discontinuities are the direct cause of the high out-of-band emission (OOBE).
To resolve this issue, we propose a fundamentally different continuous-time waveform, termed stepped frequency division multiplexing (SFDM). Unlike conventional approaches that allow continuous frequency variation, SFDM freezes the instantaneous frequency at the midpoint of the underlying chirp trajectory within each Nyquist sampling interval. This design yields a complex envelope that is strictly continuous over the entire symbol duration while preserving exact sample-wise equivalence with discrete AFDM. A unified spectral analysis reveals that the superior OOBE performance of SFDM stems from the absence of internal jump discontinuities, which otherwise dominate the far-out spectral roll-off. Numerical results confirm that SFDM consistently achieves significantly lower OOBE across a wide range of chirp rates.
\end{abstract}


\input{sec1_intro}
\input{sec2_system_model}
\input{sec3_oobe_analysis}
\input{sec4_simulation}
\input{sec5_conclusion}

\input{appendix_proofs}

\bibliographystyle{IEEEtran}
\bibliography{refs}

\end{document}

%% file: sec1_intro.tex
\section{Introduction}

Affine frequency division multiplexing (AFDM), first proposed by Bemani et al. \cite{Bemani2023AFDM}, has emerged as a promising multicarrier waveform for challenging communication scenarios, particularly in doubly dispersive channels and integrated sensing and communication (ISAC) systems \cite{cao2025agile,11173628,shao2026embodied}. Its chirp-based structure enables robust performance under high mobility and wideband conditions, distinguishing it from classical orthogonal frequency division multiplexing (OFDM). 

Despite its growing popularity, most existing studies on AFDM have adopted a purely discrete-time perspective. The waveform is typically defined through the inverse discrete affine Fourier transform (IDAFT) \cite{Bemani2021AFDM,Bemani2023AFDM,10769778}, which specifies the signal only at a set of $N$ Nyquist sampling instants. This discrete-centric approach, while convenient, leaves a fundamental question unanswered: \emph{what continuous-time waveform is actually transmitted between these samples?} However, the mapping from discrete AFDM samples to a continuous-time waveform is not unique, and the resulting realization directly affects spectral properties, hardware implementation, and regulatory compliance \cite{shao2021federated}.

Bemani et al. made the first attempt to bridge this gap \cite{Bemani2024ISAC}, where they constructed a continuous-time AFDM waveform using a frequency-wrapping mechanism that folds the instantaneous frequency of each chirp subcarrier into the Nyquist band. To ensure that the resulting continuous-time waveform exactly matches the original IDAFT samples at the sampling instants, an additional phase correction is applied at each wrapping point. This approach successfully achieves sample-wise equivalence with the discrete AFDM definition.

However, this construction has an inherent drawback. The phase correction, while preserving the sampled values, introduces abrupt changes in the instantaneous phase and complex envelope at the wrapping boundaries. As a result, the waveform is only piecewise continuous, and exhibits jump discontinuities for most chirp parameters. 
Discontinuities in the complex envelope are known to cause spectral splatter and elevated out-of-band emission (OOBE), degrading spectral containment and potentially violating spectrum mask requirements. Intriguingly, this finding also provides a theoretical explanation for the high OOBE levels previously observed in \cite{11173628}, an issue that had not been fully understood.

In this paper, we propose a different continuous-time waveform, termed stepped frequency division multiplexing (SFDM), that resolves this discontinuity issue while preserving compatibility with the original discrete AFDM framework. Instead of allowing a continuously varying wrapped chirp trajectory, SFDM uses a piecewise-constant instantaneous frequency, where the value on each Nyquist interval is chosen as the midpoint value of the underlying chirp. This yields a continuous complex envelope for arbitrary chirp parameters while exactly reproducing the original IDAFT samples at the Nyquist instants.

It is important to clarify what SFDM is and is not. Although the term {stepped frequency} appears in radar literature, those designs typically vary frequency on a pulse-by-pulse or symbol-by-symbol basis for range resolution enhancement \cite{Nguyen2016SteppedRadar,Schweizer2018SteppedCarrierOFDM,Lee2024SteppedCarrierISAC}. In contrast, SFDM operates on a much finer granularity at the Nyquist-interval level, and is inherently a multicarrier waveform. Our construction also differs from recent studies that explore continuous-time AFDM through pulse shaping or hardware impairment modeling \cite{Mirabella2026CTAFDM}; those works address complementary aspects without directly tackling the discontinuity problem we identify. SFDM is an intrinsic continuous-time waveform that modifies only the inter-sample trajectory, leaving the discrete sample values untouched.

The main contributions of this paper are as follows:
\begin{itemize}[leftmargin=0.5cm]
\item We identify and characterize an inherent discontinuity in the existing continuous-time AFDM waveform, showing that its complex envelope exhibits jumps at wrapping boundaries for generic chirp parameters, which directly explains the previously observed OOBE issue.
\item We propose SFDM, a new continuous-time waveform that achieves both strict complex-envelope continuity and exact sample-wise equivalence to discrete AFDM, without imposing any restrictions on the chirp parameter.
\item We develop a unified spectral analysis for both the conventional piecewise-continuous waveform and the proposed SFDM, revealing the fundamental mechanism behind their different OOBE behaviors.
\end{itemize}

Numerical results validate that SFDM consistently achieves lower OOBE across a wide range of chirp rates.

%% file: sec2_system_model.tex
\section{System Model}
\label{sec:system_model}

We begin by reviewing the conventional piecewise-cont\-inuous AFDM (PC-AFDM) and analyzing its inherent discontinuity, followed by the introduction of our SFDM waveform.

\subsection{Conventional PC-AFDM Waveform}
\label{subsec:pc_afdm}

Consider an AFDM block comprising $N$ subcarriers distributed over a bandwidth $B$. The block duration is defined as $T=N/B$. Under Nyquist sampling, the discrete sampling instants are given by $t_n=n/B$, for $n=0,\dots,N-1$.

For the $m$-th subcarrier, the unwrapped linear instantaneous frequency is given by $f_m^{(\mathrm{raw})}(t)=Kt+m/T$, where $K$ is the continuous-time chirp rate. To restrict the baseband frequency within $[0,B)$ (consistent with the multicarrier indices $m=0,1,\dots,N-1$), the conventional PC-AFDM construction utilizes a wrapped frequency approach, yielding the corresponding phase trajectory
\begin{equation}
\phi_m^{(\mathrm{pc})}(t)
=
\frac{K}{2}t^2+\frac{m}{T}t-q_m^{(\mathrm{pc})}(t)Bt,
\label{eq:pc_phase_def}
\end{equation}
where $q_m^{(\mathrm{pc})}(t)=\left\lfloor (Kt+m/T)/B \right\rfloor$ is the integer wrap-count. The conventional PC-AFDM waveform is formulated as
\begin{equation}
s^{(\mathrm{pc})}(t)
=
\frac{1}{\sqrt{N}}
\sum_{m=0}^{N-1}
x[m]\,
e^{j2\pi c_2 m^2}\,
e^{j2\pi\phi_m^{(\mathrm{pc})}(t)},
\quad 0\le t<T,
\label{eq:pc_tx_signal}
\end{equation}
where $x[m]$ denotes the data symbols and $c_2$ is the discrete AFDM chirp parameter.

At the sampling instants $t_n$, the phase term in \eqref{eq:pc_phase_def} exactly reproduces the standard discrete-time IDAFT basis \cite{Bemani2024ISAC}:
\begin{equation}
e^{j2\pi\phi_m^{(\mathrm{pc})}(t_n)}
=
e^{j2\pi\left(c_1 n^2+\frac{mn}{N}\right)},
\quad
c_1=\frac{K}{2B^2}.
\label{eq:pc_sample_equivalence}
\end{equation}
Throughout this paper, we consider nonnegative chirp rates, i.e., $c_1\ge0$ and equivalently $K\ge0$.
This equivalence holds because $Bt_n = n \in \mathbb{Z}$, causing the exponential of the wrapping correction term to vanish, i.e., $\exp\{-j2\pi q_m^{(\mathrm{pc})}(t_n) n\} = 1$.
However, while the correction term $-q_m^{(\mathrm{pc})}(t)Bt$ guarantees discrete equivalence at the sampling instants, its integer step changes introduce jump discontinuities in both the instantaneous phase $\phi_m^{(\mathrm{pc})}(t)$ and the transmitted complex envelope $e^{j2\pi\phi_m^{(\mathrm{pc})}(t)}$ between samples.


\begin{prop}
\label{prop:pc_continuity_condition}
The PC-AFDM waveform has a continuous complex envelope across all subcarriers
over the block interval $[0,T)$ if and only if one of the following two conditions holds:
\begin{itemize}
    \item[(i)] No internal wrapping occurs for any subcarrier, i.e.,    
    $K\le \frac{B^2}{N^2}$,    
    or equivalently, $c_1\le \frac{1}{2N^2}$, $\alpha \triangleq c_1N \le \frac{1}{2N}$.
    \item[(ii)] Internal wrapping may occur, but
    $ c_1=\frac{1}{2kN}$, $k\in\mathbb{Z}_{>0}$, 
    or equivalently, $\alpha=\frac{1}{2k}$.
\end{itemize}
\end{prop}

\begin{NewProof}
[Sketch]
For the no-wrapping regime, note that $f_m^{\mathrm{raw}}(t)=Kt+\frac{m}{T}$ is increasing in $t$, so the most stringent case is attained at $m=N-1$ and $t\to T^-$. Hence no internal wrapping occurs for any subcarrier if and only if $KT+\frac{N-1}{T}\le B$. Using $T=N/B$, this becomes $K\le \frac{B^2}{N^2}$, or equivalently $c_1\le \frac{1}{2N^2}$ and $\alpha=c_1N\le \frac{1}{2N}$. In this case, there is no internal wrapping boundary in $[0,T)$, and the complex envelope is continuous over the whole block.

Now consider the case where internal wrapping occurs. A discontinuity can only arise at an internal wrapping boundary $t=t_b$, where $q_m^{(\mathrm{pc})}(t)$ jumps by one. Since the wrapping correction contributes $-q_m^{(\mathrm{pc})}(t)Bt$ to $\phi_m^{(\mathrm{pc})}(t)$, such a jump changes the phase by $-Bt_b$. Therefore, $e^{j2\pi \phi_m^{(\mathrm{pc})}(t)}$ remains continuous at $t_b$ if and only if $Bt_b\in\mathbb{Z}$.

The wrapping boundaries satisfy $Kt_b+\frac{m}{T}=qB$ for some $q\in\mathbb{Z}$, so $t_b=\frac{qB-\frac{m}{T}}{K}$. Hence the continuity condition becomes $Bt_b=\frac{qB^2-\frac{mB}{T}}{K}\in\mathbb{Z}$. Substituting $T=N/B$ and $K=2B^2c_1$ yields $\frac{q-\frac{m}{N}}{2c_1}\in\mathbb{Z}$.

Requiring this to hold for every subcarrier and every admissible internal wrapping boundary is equivalent to $2c_1N=\frac{1}{k}$ for some $k\in\mathbb{Z}_{>0}$, namely $c_1=\frac{1}{2kN}$, or equivalently $\alpha=\frac{1}{2k}$. Combining this wrapping case with the no-wrapping case proves the proposition.
\end{NewProof}

Proposition \ref{prop:pc_continuity_condition} shows that waveform continuity holds only for a restrictive set of chirp parameters; for generic parameters, the PC-AFDM waveform suffers from jump discontinuities. Such discontinuities are known to significantly increase OOBE and are highly undesirable in practice \cite{tektronix_radar_primer}.

\subsection{The SFDM Waveform}
\label{subsec:midpoint_stepped}

We next introduce SFDM, which eliminate the phase discontinuities while preserving full parameter flexibility and exact discrete-time equivalence.

\begin{defi}
The SFDM waveform is defined as
\begin{equation*}
s^{(\mathrm{step})}(t)
\!=\!
\frac{1}{\sqrt{N}}\!
\sum_{m=0}^{N-1}
x[m]\,
e^{j2\pi c_2 m^2}\,
e^{j2\pi\phi_m^{(\mathrm{step})}(t)},~ 0\le t<T,
\end{equation*}
where the continuous phase is accumulated from a step-frequency trajectory:
\begin{equation}
\phi_m^{(\mathrm{step})}(t)=\int_0^t f_m^{(\mathrm{step})}(\tau)\,d\tau.
\label{eq:step_phase_integral}
\end{equation}
For $t\in[n/B,(n+1)/B)$, the instantaneous frequency is frozen at the underlying chirp's midpoint to remain bounded within $[0,B)$:
\begin{equation*}
f_m^{(\mathrm{step})}(t)
=
K\frac{n+\tfrac{1}{2}}{B} + \frac{m}{T} - B \left\lfloor \frac{K(n+\tfrac{1}{2})/B+m/T}{B} \right\rfloor.
\end{equation*}
\end{defi}

To avoid numerical integration, the continuous phase is generated recursively. Let $f_{m,n}^{(\mathrm{step})}$ be the constant frequency value on the $n$-th interval, the phase updates at interval boundaries as $\phi_m^{(\mathrm{step})}\left(\frac{n+1}{B}\right) = \phi_m^{(\mathrm{step})}\left(\frac{n}{B}\right) + \frac{1}{B}f_{m,n}^{(\mathrm{step})}$, with linear intra-interval evolution $\phi_m^{(\mathrm{step})}(t) = \phi_m^{(\mathrm{step})}\left(\frac{n}{B}\right) + f_{m,n}^{(\mathrm{step})}\left(t - \frac{n}{B}\right)$.

\begin{prop}
\label{prop:step_sample_equivalence}
The sampled SFDM subcarrier basis coincides exactly with the standard discrete-time AFDM/IDAFT basis for arbitrary $c_1$, namely
\begin{equation}
e^{j2\pi\phi_m^{(\mathrm{step})}(t_n)}
=
e^{j2\pi\left(c_1 n^2+\frac{mn}{N}\right)}.
\label{eq:step_sample_equivalence}
\end{equation}
\end{prop}

\begin{NewProof}
[Sketch]
At the sampling instant $t_n=n/B$,
$\phi_m^{(\mathrm{step})}(t_n)
=
\sum_{r=0}^{n-1}\frac{1}{B}\left(
K\frac{r+\tfrac{1}{2}}{B}+\frac{m}{T}-Bq_m^{(\mathrm{step})}[r]
\right)$.
Using $\sum_{r=0}^{n-1}(r+\tfrac{1}{2})=n^2/2$ and $T=N/B$, we obtain
$\phi_m^{(\mathrm{step})}(t_n)
=
\frac{K}{2B^2}n^2+\frac{mn}{N}-\sum_{r=0}^{n-1}q_m^{(\mathrm{step})}[r]$.
Since $\sum_{r=0}^{n-1}q_m^{(\mathrm{step})}[r]\in\mathbb{Z}$ and $c_1=K/(2B^2)$, exponentiation yields
$e^{j2\pi\phi_m^{(\mathrm{step})}(t_n)} = e^{j2\pi\left(c_1n^2+\frac{mn}{N}\right)}$.
\end{NewProof}

Let the elements of the sampled subcarrier matrix $\bm{A}_{\xi}$ for $\xi \in \{\mathrm{pc}, \mathrm{step}, \mathrm{IDAFT}\}$ be $[\bm{A}_{\xi}]_{n,m} \triangleq \frac{1}{\sqrt{N}} \exp\{j2\pi(c_2 m^2 + \phi_m^{(\xi)}(t_n))\}$. Proposition \ref{prop:step_sample_equivalence} explicitly implies that $\bm{A}_{\mathrm{step}} \equiv \bm{A}_{\mathrm{IDAFT}}$. Consequently, the SFDM construction inherently satisfies $\bm{A}_{\mathrm{step}}^{H}\bm{A}_{\mathrm{step}}=\bm{I}$, ensuring perfect discrete-time subcarrier orthogonality without any transceiver modifications.

\begin{rem}
The midpoint choice is motivated by symmetry: it yields a centered piecewise-constant representative over each Nyquist interval, unlike left- or right-endpoint choices, which are locally one-sided.
\end{rem}


\begin{rem}
Since both the PC-AFDM and SFDM constructions reproduce the same Nyquist-rate AFDM/IDAFT samples, they are fully compatible with the standard discrete-time AFDM transceiver. Therefore, the difference studied in this paper lies entirely in the continuous-time inter-sample trajectory and its spectral consequences, rather than in the discrete-time receiver structure.
\end{rem}

%% file: sec3_oobe_analysis.tex
\section{Continuous-Time Spectral Characterization}
\label{sec:oobe_analysis}
Since both waveforms share the same Nyquist-rate samples, any spectral difference arises solely from their inter-sample trajectories. To isolate this effect, we analyze the continuous-time spectrum over one block $0\le t<T$.


\label{subsec:unified_spectral_metrics}

Denote by $S^{(\xi)}(f)
\triangleq
\int_0^T s^{(\xi)}(t)e^{-j2\pi ft}\,dt$ the continuous-time spectrum over one block. From Section~\ref{sec:system_model}, we have
\begin{equation}
\label{eq:continuous_tx_signal}
s^{(\xi)}(t)
=
\frac{1}{\sqrt N}\sum_{m=0}^{N-1}
x[m]\,e^{j2\pi c_2 m^2}\,g_m^{(\xi)}(t),
\quad 0\le t<T,
\end{equation}
where $g_m^{(\xi)}(t)=e^{j2\pi\phi_m^{(\xi)}(t)}$ is the $m$-th subcarrier.

\begin{defi}[OOBE]
\label{def:psd_oobe}
Assume that $\{x[m]\}$ are independent, zero-mean, and unit-variance. Let
\begin{equation}
\label{eq:average_psd_definition}
\Phi_{\xi}(f)
\triangleq
\mathbb{E}\!\left[\left|S^{(\xi)}(f)\right|^2\right]
=
\frac{1}{N}\sum_{m=0}^{N-1}\left|G_m^{(\xi)}(f)\right|^2
\end{equation}
denote the average energy spectral density (ESD), where $G_m^{(\xi)}(f)
\triangleq
\int_0^T g_m^{(\xi)}(t)e^{-j2\pi ft}\,dt$ is the spectrum of the $m$-th subcarrier.

With the in-band region defined as $[0,B)$, the average out-of-band energy is
\begin{equation}
\label{eq:average_oobe_energy}
\overline P_{\mathrm{OOBE}}^{(\xi)}
\triangleq
\int_{\mathbb{R}\setminus[0,B)} \Phi_{\xi}(f)\,df.
\end{equation}
Moreover, by Parseval's identity, the total average energy satisfies $\int_{-\infty}^{\infty}\Phi_{\xi}(f)\,df
=
\mathbb{E}\!\left[\int_0^T |s^{(\xi)}(t)|^2\,dt\right]
=
T$,
and hence the normalized OOBE ratio is defined as
\begin{equation}
\label{eq:normalized_oobe_ratio}
\eta_{\mathrm{OOBE}}^{(\xi)}
\triangleq
\frac{\overline P_{\mathrm{OOBE}}^{(\xi)}}{T}.
\end{equation}
\end{defi}

Thus, the spectral difference between the two waveforms is entirely determined by the subcarrier spectra $\{G_m^{(\xi)}(f)\}$.

\begin{prop}[Exact subcarrier spectral representations]
\label{prop:subcarrier_spectra_closed_form}
The spectrum of SFDM admits the exact sinc-sum form
\begin{equation*}
\begin{aligned}
G_m^{(\mathrm{step})}(f)
&=
\frac{1}{B}
\sum_{n=0}^{N-1}
\exp\!\left[
j2\pi
\left(
\phi_{m,n}^{(0)}-\frac{fn}{B}
\right)
\right] \\
&\quad \times
\exp\!\left[
j\pi\frac{f_{m,n}^{(\mathrm{step})}-f}{B}
\right]
\mathrm{sinc}\!\left(
\frac{f_{m,n}^{(\mathrm{step})}-f}{B}
\right),
\end{aligned}
\end{equation*}
where $\phi_{m,n}^{(0)}\triangleq \phi_m^{(\mathrm{step})}(n/B)$ and $\mathrm{sinc}(x)\triangleq \sin(\pi x)/(\pi x)$.

For the PC-AFDM waveform, let $0=t_{m,0}<t_{m,1}<\cdots<t_{m,J_m}<t_{m,J_m+1}=T$ be the ordered internal wrapping boundaries in $[0,T]$, and let $q_{m,j}$ denote the constant wrap-count on $[t_{m,j},t_{m,j+1})$. Then
\begin{equation}
\label{eq:pc_subcarrier_spectrum_prop_integral}
\begin{aligned}
G_m^{(\mathrm{pc})}(f)
&=
\sum_{j=0}^{J_m}
\int_{t_{m,j}}^{t_{m,j+1}}
\exp\!\Bigg[
j2\pi
\Bigg(
\frac{K}{2}t^2 \\
&\qquad\qquad +
\left(
\frac{m}{T}-q_{m,j}B-f
\right)t
\Bigg)
\Bigg]dt.
\end{aligned}
\end{equation}
Equivalently, for $K>0$,
\begin{equation}
\label{eq:pc_subcarrier_spectrum_prop_fresnel}
\begin{aligned}
G_m^{(\mathrm{pc})}(f)
&=
\frac{1}{\sqrt{2K}}
\sum_{j=0}^{J_m}
\exp\!\left(
-j\pi\frac{\beta_{m,j}^2(f)}{K}
\right) \\
&\quad \times
\left[
\mathcal{F}_{\mathrm{Fr}}\!\bigl(u_{m,j+1}(f)\bigr)
-
\mathcal{F}_{\mathrm{Fr}}\!\bigl(u_{m,j}(f)\bigr)
\right],
\end{aligned}
\end{equation}
where $\beta_{m,j}(f)
\triangleq
\frac{m}{T}-q_{m,j}B-f$, $u_{m,j}(f)
\triangleq
\sqrt{2K}
\left(
t_{m,j}+\frac{\beta_{m,j}(f)}{K}
\right)$, and the standard Fresnel integral is defined as $\mathcal{F}_{\mathrm{Fr}}(u)
\triangleq
\int_0^u e^{j\pi v^2/2}\,dv$.
\end{prop}

\begin{NewProof}
To conserve space, the proof is presented in our technical report \cite{SFDMtech}.
\end{NewProof}

\begin{rem}
As can be seen, the SFDM spectrum is a coherent sum of sinc-shaped terms associated with constant-frequency intervals, whereas the PC-AFDM spectrum is a coherent sum of segment-dependent Fresnel-type terms induced by wrapping boundaries.
Consequently, although both spectra are exact, only the proposed waveform admits a compact representation directly tied to intervalwise constant-frequency segments.
\end{rem}

To compare the two spectra analytically, we distinguish two regimes. In the small-$\alpha$ regime, the two trajectories are locally close on intervals without internal wrapping. In the far-out regime, their asymptotic tails are governed by different regularity structures, namely internal jump discontinuities for PC-AFDM versus internal continuity for SFDM.

\begin{prop}[Small-$\alpha$ local phase boundedness]
\label{prop:small_alpha_bound}
Consider an interval $I_n=[n/B,(n+1)/B)$ in which no frequency wrapping occurs for the $m$-th PC subcarrier, i.e., $0 \le Kt + m/T < B$ for all $t \in I_n$. Then, for all $t\in I_n$,
\begin{equation}
\label{eq:local_phase_difference_bound_alpha}
\left|
\phi_m^{(\mathrm{pc})}(t)
-
\phi_m^{(\mathrm{step})}(t)
-
C_{m,n}
\right|
\le
\frac{\alpha}{4N},
\end{equation}
where $C_{m,n}$ is an interval-dependent constant.
\end{prop}

\begin{NewProof}
[Sketch]
Let $t_c=(n+\tfrac12)/B$ denote the midpoint of $I_n$. Since no wrapping occurs inside $I_n$, the PC instantaneous frequency on this interval is affine, and can be written as $a_n+K(t-t_c)$, where $a_n$ is its value at $t_c$. Over the same interval, the SFDM waveform uses the constant frequency $a_n$. Therefore, $\frac{d}{dt}\big(\phi_m^{(\mathrm{pc})}(t)-\phi_m^{(\mathrm{step})}(t)\big)=K(t-t_c)$, which yields $\phi_m^{(\mathrm{pc})}(t)-\phi_m^{(\mathrm{step})}(t)=\frac{K}{2}(t-t_c)^2+C_{m,n}$ for some interval-dependent constant $C_{m,n}$.

Since $|t-t_c|\le \frac{1}{2B}$ for all $t\in I_n$, it follows that $\left|\phi_m^{(\mathrm{pc})}(t)-\phi_m^{(\mathrm{step})}(t)-C_{m,n}\right|\le \frac{K}{8B^2}$. Finally, using $c_1=\frac{K}{2B^2}$ and $\alpha=c_1N$, we obtain $\frac{K}{8B^2}=\frac{c_1}{4}=\frac{\alpha}{4N}$, which proves \eqref{eq:local_phase_difference_bound_alpha}.
\end{NewProof}

Proposition~\ref{prop:small_alpha_bound} locally bounds the phase difference by a small quadratic perturbation on intervals without internal wrapping. This explains why the two spectra are close when $\alpha$ is small. However, this local closeness does not by itself imply a global ordering of the full OOBE.

As $\alpha$ increases, internal wrapping events become increasingly important. SFDM remains continuous inside $[0,T)$, whereas PC-AFDM generically develops internal jump discontinuities in its complex envelope. The far-out spectral behavior is therefore governed not merely by local phase mismatch, but by this difference in waveform regularity. We formalize this mechanism through a high-frequency expansion.

\begin{prop}[High-frequency expansion and jump-induced leading term]
\label{prop:high_frequency_expansion}
Assume that $g_m^{(\mathrm{pc})}(t)$ is piecewise $C^2$ on $[0,T)$, i.e., there exist finitely many points
$
0<t_{m,1}<\cdots<t_{m,J_m}<T
$
such that $g_m^{(\mathrm{pc})}(t)\in C^2$ on each open subinterval determined by these points, with internal jump sizes
$
\Delta g_{m,j}
\triangleq
g_m^{(\mathrm{pc})}(t_{m,j}^{+})
-
g_m^{(\mathrm{pc})}(t_{m,j}^{-})$, $j=1, \dots,J_m$,
where $\{t_{m,j}\}_{j=1}^{J_m}$ are the internal wrapping boundaries
defined in Proposition~\ref{prop:subcarrier_spectra_closed_form}.
Assume also that $g_m^{(\mathrm{step})}(t)$ is continuous on $[0,T]$ and
$C^2$ on each Nyquist interval. Then, as $|f|\to\infty$,
\begin{equation}
\begin{aligned}
G_m^{(\mathrm{pc})}(f)
&=
\frac{1}{j2\pi f}
\Bigg[
g_m^{(\mathrm{pc})}(0^{+})
-
g_m^{(\mathrm{pc})}(T^{-})e^{-j2\pi fT} \\
&\qquad
+
\sum_{j=1}^{J_m}
\Delta g_{m,j}e^{-j2\pi f t_{m,j}}
\Bigg]
+
O(f^{-2}),
\end{aligned}
\end{equation}
whereas
\begin{equation}
\label{eq:step_high_frequency_expansion}
\begin{aligned}
G_m^{(\mathrm{step})}(f)
&= \frac{g_m^{(\mathrm{step})}(0^{+})}{j2\pi f} \\
&\quad - \frac{g_m^{(\mathrm{step})}(T^{-})e^{-j2\pi fT}}{j2\pi f}
+ O(f^{-2}).
\end{aligned}
\end{equation}
\end{prop}

\begin{NewProof}
[Sketch]
Partition $[0,T]$ by the internal jump points $0=t_{m,0}<t_{m,1}<\cdots<t_{m,J_m}<t_{m,J_m+1}=T$. On each interval $(t_{m,j},t_{m,j+1})$, $g_m^{(\mathrm{pc})}(t)$ is $C^2$. Applying integration by parts and using the piecewise smoothness of its derivative gives
\[
\begin{aligned}
\int_{t_{m,j}}^{t_{m,j+1}} g_m^{(\mathrm{pc})}(t)e^{-j2\pi ft}\,dt
&=
\frac{1}{j2\pi f} \Big[ g_m^{(\mathrm{pc})}(t_{m,j}^{+})e^{-j2\pi f t_{m,j}} \\
&\quad - g_m^{(\mathrm{pc})}(t_{m,j+1}^{-})e^{-j2\pi f t_{m,j+1}} \Big] \\
&\quad + O(f^{-2}) .
\end{aligned}\]
Summing over $j$, the interior endpoint terms telescope, and only the endpoint contributions plus the jump mismatches remain, yielding the stated $f^{-1}$ expansion for $G_m^{(\mathrm{pc})}(f)$.

For $g_m^{(\mathrm{step})}(t)$, the same argument applies, but continuity on $[0,T)$ removes all internal jump terms. Hence only the endpoint contribution remains at order $f^{-1}$, which gives \eqref{eq:step_high_frequency_expansion}.

\end{NewProof}

\begin{rem}
The jump-induced leading term in $G_m^{(\mathrm{pc})}(f)$ disappears under the discrete
continuity condition of Proposition~\ref{prop:pc_continuity_condition}. In this case, every internal wrapping boundary
$t_{m,j}$ satisfies $Bt_{m,j}\in\mathbb{Z}$, and hence
$
\Delta g_{m,j}
=
g_m^{(\mathrm{pc})}(t_{m,j}^{-})
\left(e^{-j2\pi Bt_{m,j}}-1\right)
=
0$.
Therefore, although internal wrapping may still occur, it does not
contribute any internal-jump term to the $1/f$ leading-order expansion.
\end{rem}

Proposition~\ref{prop:high_frequency_expansion} shows that, as $|f|\to\infty$, the $f^{-1}$-order leading term of the generic PC-AFDM subcarrier contains explicit contributions from internal jump discontinuities, whereas that of the SFDM subcarrier involves no such internal jump contribution. Therefore, the far-out asymptotic difference between the two waveforms is fundamentally tied to the presence or absence of internal discontinuities.

For the PC-AFDM waveform, each internal jump satisfies
\begin{equation}
\label{eq:jump_magnitude_pc}
\begin{aligned}
\Delta g_{m,j}
&=
g_m^{(\mathrm{pc})}(t_{m,j}^{+})
-
g_m^{(\mathrm{pc})}(t_{m,j}^{-}) \\
&=
g_m^{(\mathrm{pc})}(t_{m,j}^{-})
\left(
e^{-j2\pi B t_{m,j}}-1
\right),
\end{aligned}
\end{equation}
and hence
\begin{equation}
\label{eq:jump_magnitude_squared_pc}
|\Delta g_{m,j}|^2
=
4\sin^2(\pi B t_{m,j}).
\end{equation}
Therefore, whenever internal wrapping occurs outside the restrictive continuity conditions of Proposition~\ref{prop:pc_continuity_condition}, the corresponding jump contribution is generically nonzero and produces an additional far-out penalty absent from SFDM.

%% file: sec4_simulation.tex
\section{Numerical Results}
\label{sec:numerical_results}

This section evaluates the continuous-time behavior and OOBE of the proposed SFDM against the conventional PC-AFDM baseline. The simulation settings are summarized in Table~\ref{tab:parameters}. A normalized bandwidth $B=1$ is adopted, and continuous-time waveforms are numerically represented with oversampling factor $L_{\mathrm{os}}$. The trajectory plots use $N=10$, while the ESD and OOBE evaluations use $N=64$. In Experiment~2, the average ESD $\Phi(f)$ is computed deterministically from \eqref{eq:average_psd_definition} via a zero-padded FFT, and the OOBE integral in \eqref{eq:average_oobe_energy} is evaluated numerically over the truncation range in Table~\ref{tab:parameters}.



\begin{table}[t]
    \centering
    \scriptsize
    \caption{Simulation Parameters}
    \label{tab:parameters}
    \setlength{\tabcolsep}{5pt}
    \renewcommand{\arraystretch}{0.95}
    \begin{tabular}{lccc}
        \toprule
        \textbf{Parameter} & \textbf{Trajectory} & \textbf{ESD} & \textbf{OOBE} \\
        \midrule
        $N$ & 10 & 64 & 64 \\
        $\alpha$ & $\{0.5,\,0.8\}$ & $\{0.5,\,0.8\}$ & $[0,1]$ \\
        $N_{\mathrm{FFT}}$ & N/A & $256{,}000$ & $256{,}000$ \\
        Freq. range & N/A & \begin{tabular}[c]{@{}c@{}}$[-3B,\,4B]$ \\ $[-3B,\,-0.5B]\cup[1.5B,\,4B]$\end{tabular} & $[-250B,\,250B)$ \\
        \bottomrule
    \end{tabular}
    
    \vspace{1mm}
    {\scriptsize Common settings: $B=1$ and oversampling factor $L_{\mathrm{os}}=500$.}
\end{table}

\subsection{Time-Frequency Trajectories and Phase Continuity}

To start with, we investigate the instantaneous frequency trajectories and continuous-time waveforms for selected subcarriers ($m \in \{0, 3\}$) under different normalized chirp rates ($\alpha = 0.5$ and $\alpha = 0.8$). 

\begin{figure}[!t]
    \centering
    \includegraphics[width=0.88\linewidth]{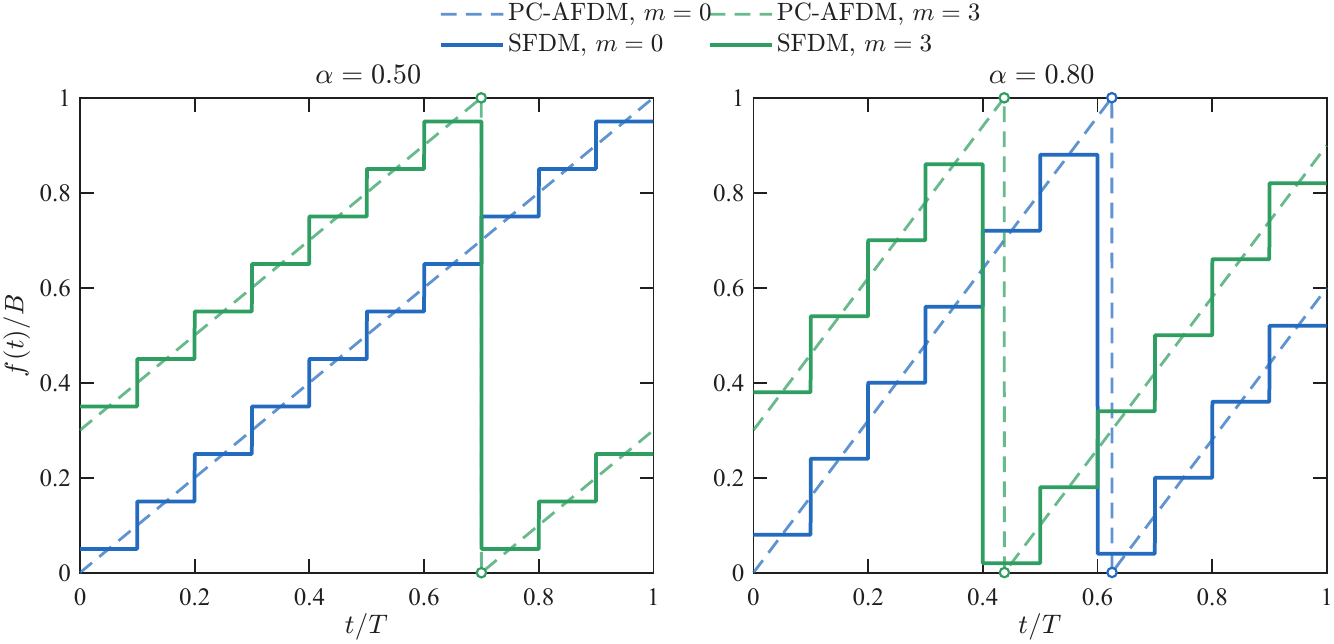}
    \caption{Instantaneous time-frequency trajectories of the PC-AFDM and SFDM waveforms for $\alpha = 0.5$ and $\alpha = 0.8$.}
    \label{fig:tf_traj}
\end{figure}

As shown in Fig. \ref{fig:tf_traj}, under these $\alpha$ settings, both waveforms experience frequency wrapping to stay within the baseband limits. The instantaneous frequency of the proposed SFDM is frozen at the midpoint frequency of the underlying linear chirp within each Nyquist sampling interval.

\begin{figure}[!t]
    \centering
    \includegraphics[width=0.65\linewidth]{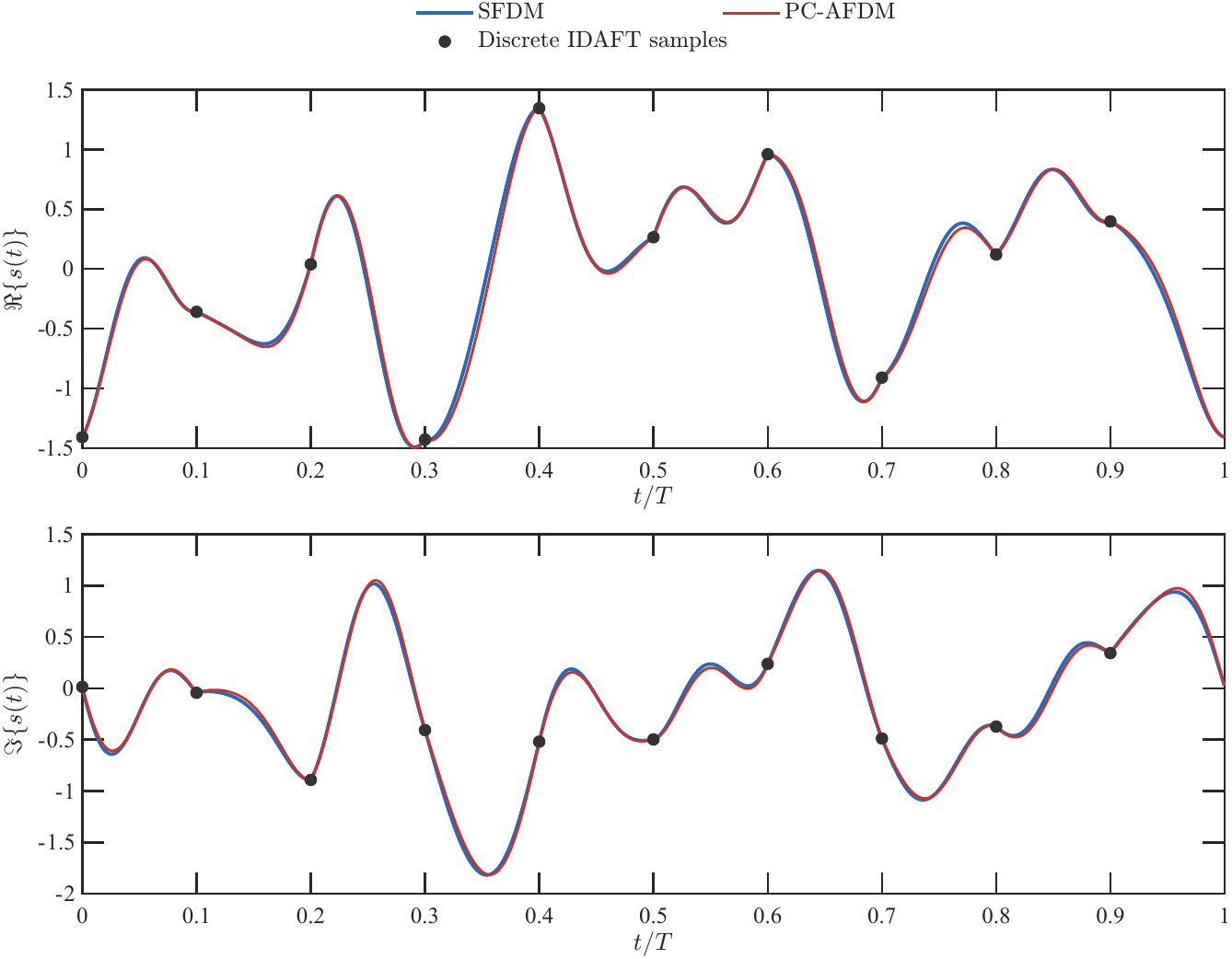}
    \caption{Real and imaginary components of the continuous-time waveforms for $\alpha = 0.5$. At this specific chirp rate, both waveforms preserve continuity.}
    \label{fig:waveform_alpha05}
\end{figure}

\begin{figure}[!t]
    \centering
    \includegraphics[width=0.65\linewidth]{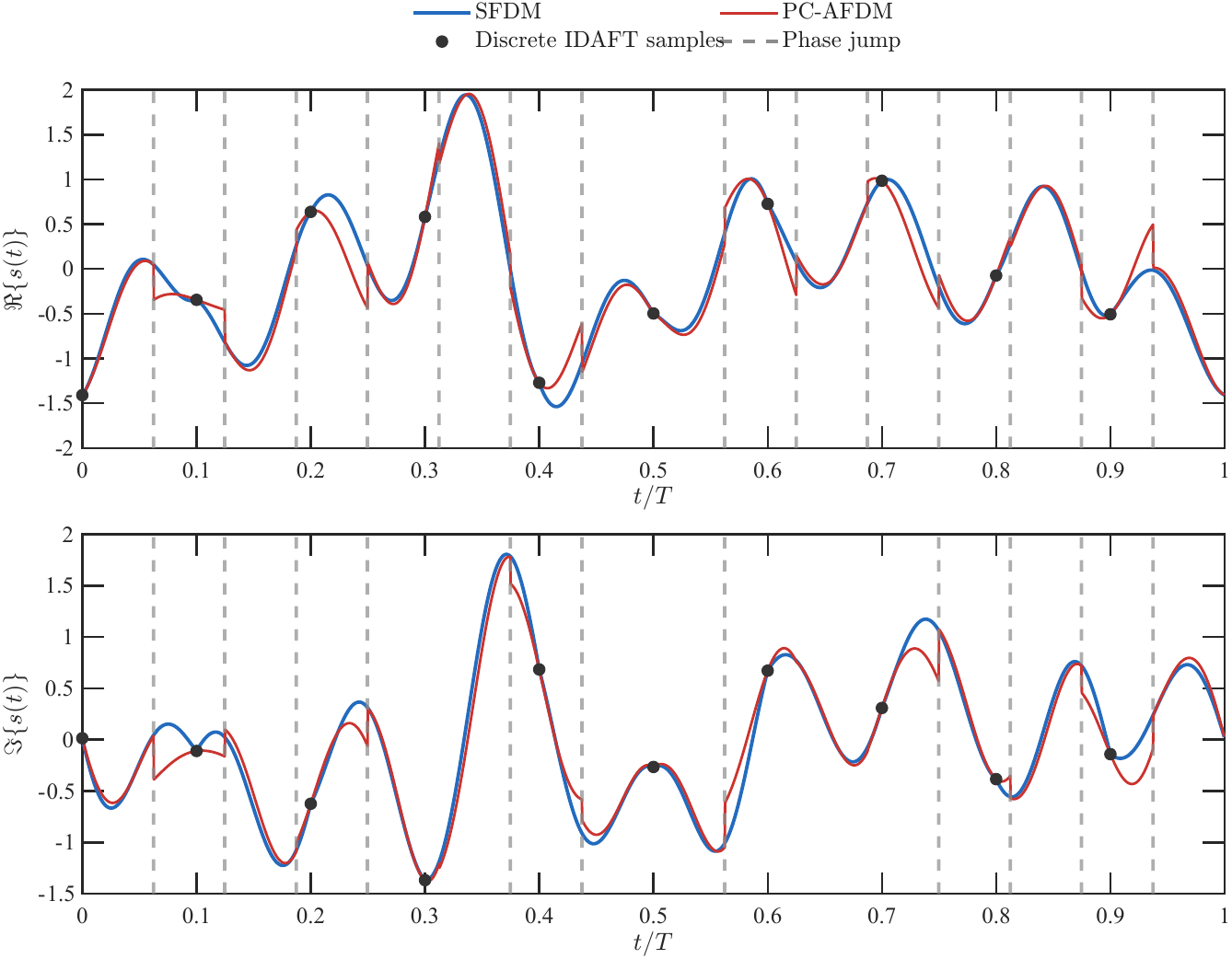}
    \caption{Real and imaginary components of the continuous-time waveforms for $\alpha = 0.8$. The PC-AFDM waveform exhibits jump discontinuities, whereas the SFDM remains continuous.}
    \label{fig:waveform_alpha08}
\end{figure}

Figs.~\ref{fig:waveform_alpha05} and \ref{fig:waveform_alpha08} illustrate the real and imaginary components of the generated continuous-time signals. At the sampling instants, both continuous-time constructions precisely intersect with the discrete IDAFT samples, confirming the exact sample-wise equivalence established in Proposition~\ref{prop:step_sample_equivalence}. 

However, their inter-sample behaviors differ significantly. For a generic chirp rate such as $\alpha = 0.8$ (as shown in Fig.~\ref{fig:waveform_alpha08}), the PC-AFDM waveform exhibits prominent jump discontinuities in its complex envelope, triggered by the internal frequency wrappings. In contrast, the SFDM waveform strictly preserves complex envelope continuity over the entire symbol duration, eliminating non-continuous phase variations regardless of the chirp rate. 

Furthermore, under the restrictive condition of $\alpha = 0.5$ (as shown in Fig.~\ref{fig:waveform_alpha05}), the phase jumps in the PC-AFDM waveform become exact integer multiples of $2\pi$, which mathematically preserves its waveform continuity and explicitly corroborates Proposition~\ref{prop:pc_continuity_condition}.

\subsection{Out-of-Band Emission Performance}

\begin{figure}[!t]
    \centering
    \includegraphics[width=0.77\linewidth]{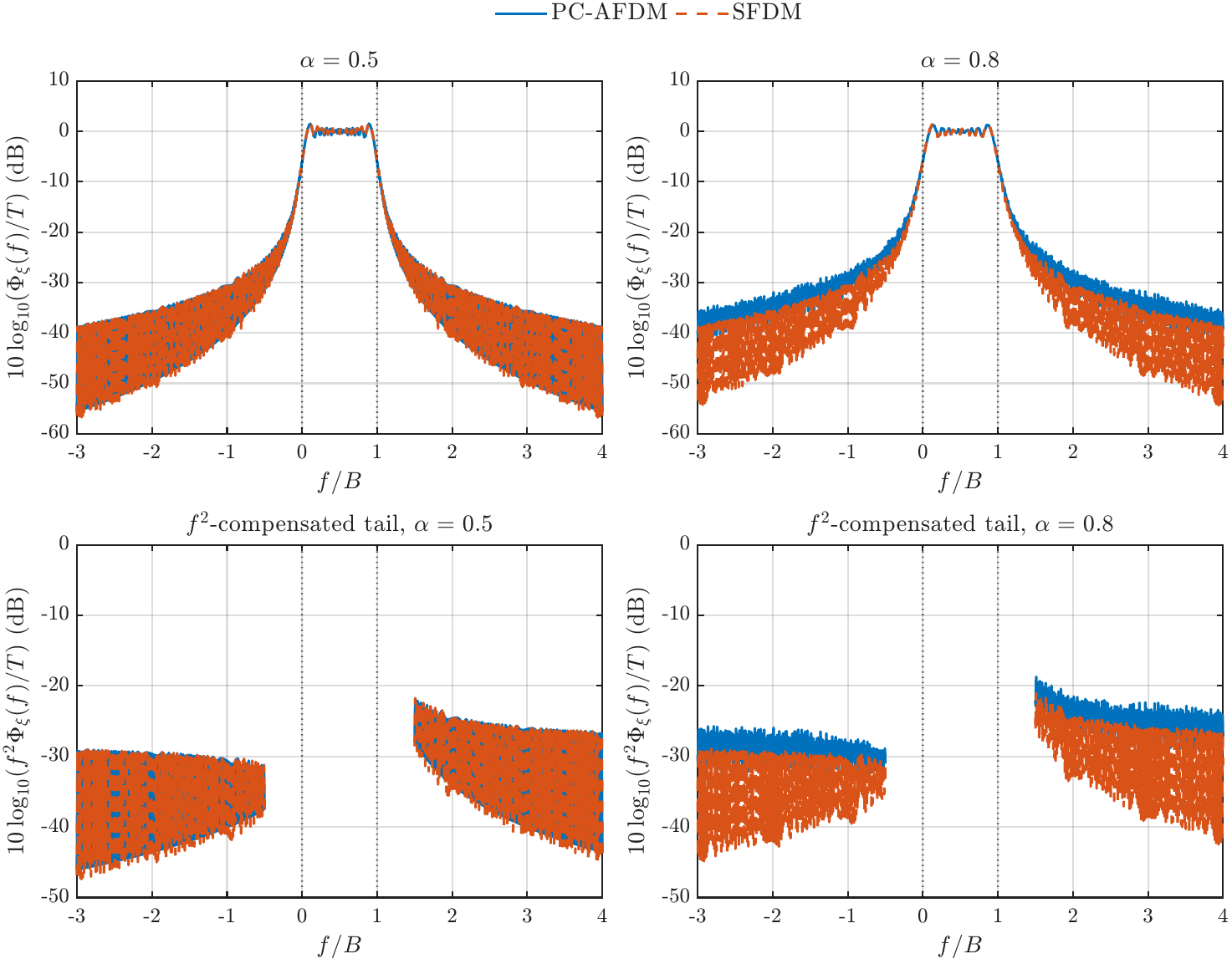}
    \caption{ESD comparison between PC-AFDM and SFDM. Top row: normalized ESD over $f/B\in[-3,4]$. Bottom row: compensated far-out ESD, shown only over $f/B\in[-3,-0.5]\cup[1.5,4]$. Left column: $\alpha=0.5$, where the PC waveform satisfies the continuity condition. Right column: $\alpha=0.8$, corresponding to a generic discontinuous PC-AFDM case.}
    \label{fig:spec_density_representative}
\end{figure}

Before evaluating the OOBE performance, we first examine representative ESD plots to visualize the spectral shapes of the two continuous-time waveforms. The top row of Fig.~\ref{fig:spec_density_representative} shows the normalized ESD over $f/B\in[-3,4]$, while the bottom row shows the compensated far-out ESD over $f/B\in[-3,-0.5]\cup[1.5,4]$, where the comparison is not visually dominated by the near-band portion around the useful band $[0,B)$.

When $\alpha=0.5$, the ESDs of PC-AFDM and SFDM are nearly indistinguishable, consistent with the disappearance of internal jump discontinuities at this special chirp rate. In contrast, when $\alpha=0.8$, PC-AFDM exhibits stronger far-out leakage, which becomes even clearer in the compensated far-out ESD plot. This agrees with Proposition~\ref{prop:high_frequency_expansion}, which shows that internal-jump contributions appear in the $f^{-1}$-order leading term of the generic PC-AFDM subcarrier but not in that of the SFDM subcarrier.

Next, we evaluate the average OOBE, $\eta_{\mathrm{OOBE}}$, as a function of the normalized chirp rate $\alpha \in [0, 1]$.

\begin{figure}[!t]
    \centering
    \includegraphics[width=0.77\linewidth]{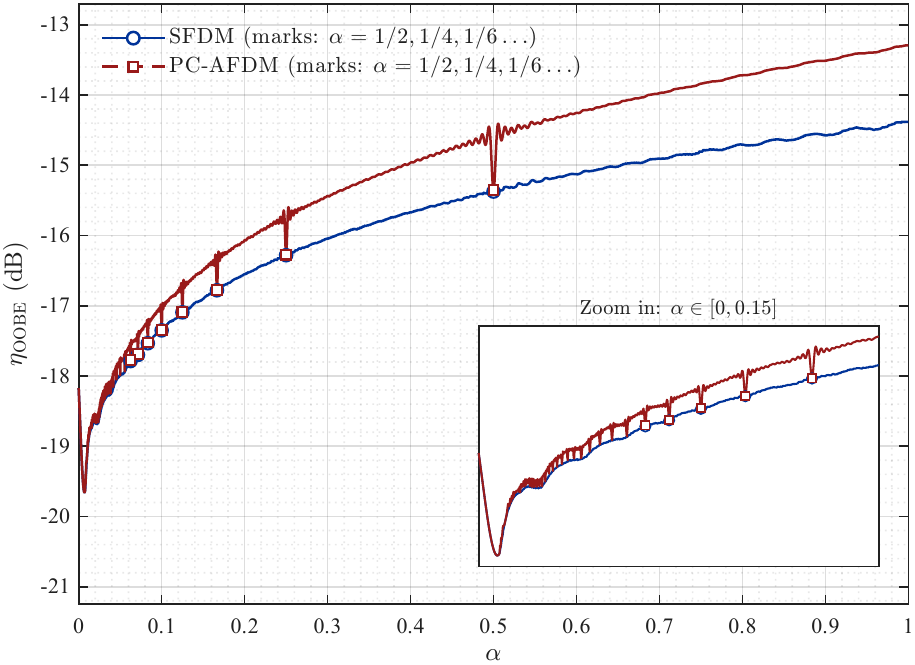}
    \caption{Average OOBE $\eta_{\mathrm{OOBE}}$ versus the normalized chirp rate $\alpha$. Marks indicate the theoretical continuity points $\alpha = 1/(2k)$ for $k=1,2,3,\ldots$.}
    \label{fig:oobe_alpha}
\end{figure}

Fig.~\ref{fig:oobe_alpha} shows that SFDM yields a considerably lower OOBE than the PC-AFDM baseline over most of the evaluated $\alpha$ range. In the small-$\alpha$ regime, e.g., $\alpha \le 0.006$, the OOBE curves of the two waveforms closely overlap. This is because no internal frequency wrapping occurs in PC-AFDM within this range, while the inter-sample phase mismatch remains small according to Proposition~\ref{prop:small_alpha_bound}. Consequently, the two constructions remain locally close on average and exhibit similar spectral behavior. As $\alpha$ increases, internal wrappings in PC-AFDM become more frequent and widespread across the multiplexing block, creating additional far-out spectral leakage and hence higher OOBE.

The numerical results also show that the OOBE values of the two waveforms nearly coincide at the discrete points $\alpha = 1/(2k)$, where $k \in \mathbb{Z}_{>0}$, e.g., $\alpha = 1/2, 1/4, 1/6$. At these special values, the phase jumps of the PC-AFDM waveform become exact integer multiples of $2\pi$, thereby preserving complex-envelope continuity. This agrees with Proposition~1. Outside these restrictive conditions, the SFDM waveform consistently avoids jump-induced spectral penalties and therefore provides lower OOBE for AFDM systems with flexible chirp parameters.

%% file: sec5_conclusion.tex
\section{Conclusion}
\label{sec:conclusion}

This paper has revealed and addressed a critical but previously overlooked flaw in continuous-time AFDM waveforms: the inherent jump discontinuities in the complex envelope that lead to high OOBE. The proposed SFDM offers a principled alternative that eliminates these discontinuities while preserving exact sample-wise equivalence. More broadly, this work establishes that for chirp-based multicarrier systems such as AFDM, the continuous-time realization is not a mere implementation detail but a fundamental design choice that directly governs spectral containment. By resolving the discontinuity issue, SFDM removes a key obstacle to the practical deployment of AFDM in spectrum-constrained applications, and opens the door to further exploration of stepped-frequency principles in next-generation waveform design.

%% file: appendix_proofs.tex
\appendices

%% file: refs.bib
@INPROCEEDINGS{Bemani2021AFDM,
  author={Bemani, Ali and Ksairi, Nassar and Kountouris, Marios},
  booktitle={IEEE ICC Workshops}, 
  title={{AFDM}: A Full Diversity Next Generation Waveform for High Mobility Communications}, 
  year={2021}
  }

@ARTICLE{Bemani2023AFDM,
  author={Bemani, Ali and Ksairi, Nassar and Kountouris, Marios},
  journal={IEEE Trans. Wireless Commun.}, 
  title={Affine Frequency Division Multiplexing for Next Generation Wireless Communications}, 
  year={2023},
  volume={22},
  number={11},
  pages={8214-8229}
  }

@ARTICLE{Bemani2024ISAC,
  author={Bemani, Ali and Ksairi, Nassar and Kountouris, Marios},
  journal={IEEE Wireless Commun. Lett.}, 
  title={Integrated Sensing and Communications With Affine Frequency Division Multiplexing}, 
  year={2024},
  volume={13},
  number={5},
  pages={1255-1259},
  doi={10.1109/LWC.2024.3367178}}

@book{Nguyen2016SteppedRadar,
  author    = {Cam Nguyen and Joongsuk Park},
  title     = {Stepped-Frequency Radar Sensors: Theory, Analysis and Design},
  publisher = {Springer},
  address   = {Cham, Switzerland},
  year      = {2016},
  doi       = {10.1007/978-3-319-12271-7},
  isbn={978-3-319-12271-7}
}

@ARTICLE{Schweizer2018SteppedCarrierOFDM,
  author={Schweizer, Benedikt and Knill, Christina and Schindler, Daniel and Waldschmidt, Christian},
  journal={IEEE Transactions on Microwave Theory and Techniques}, 
  title={Stepped-Carrier {OFDM}-Radar Processing Scheme to Retrieve High-Resolution Range-Velocity Profile at Low Sampling Rate}, 
  year={2018},
  volume={66},
  number={3},
  pages={1610-1618},
  doi={10.1109/TMTT.2017.2751463}}

@INPROCEEDINGS{Lee2024SteppedCarrierISAC,
  author={Lee, Kyung In and Mung Shin, Jae and Kim, Dong In and Won Choi, Kae},
  booktitle={IEEE SPAWC}, 
  title={An Experimental Proof of Concept for {OFDM}-based {ISAC} System Utilizing Stepped-Carrier and {TDM MIMO} scheme}, 
  year={2024},
  volume={},
  number={},
  pages={601-605}
  }

@misc{tektronix_radar_primer,
  author       = {{Tektronix, Inc.}},
  title        = {Understanding Radar Signals Using Real-Time Spectrum Analysis},
  howpublished = {Application Primer},
  year         = {2018}
}

@article{Mirabella2026CTAFDM,
  title={Continuous-Time Analysis of {AFDM}: Pulse-Shaping, Fundamental Bounds and Impact of Hardware Impairments},
  author={Michele Mirabella and Hyeon Seok Rou and Pasquale Di Viesti and Giuseppe Thadeu Freitas de Abreu and Giorgio Matteo Vitetta},
  journal={arXiv preprint arXiv:2602.20909},
  year={2026}
}

@ARTICLE{10769778,
  author={Rou, Hyeon Seok and de Abreu, Giuseppe Thadeu Freitas and Choi, Junil and González G., David and Kountouris, Marios and Guan, Yong Liang and Gonsa, Osvaldo},
  journal={IEEE Signal Processing Magazine}, 
  title={From Orthogonal Time-Frequency Space to Affine Frequency-Division Multiplexing: A comparative study of next-generation waveforms for integrated sensing and communications in doubly dispersive channels}, 
  year={2024},
  volume={41},
  number={5},
  pages={71-86},
  doi={10.1109/MSP.2024.3422653}}

@article{cao2025agile,
  title={Agile Affine Frequency Division Multiplexing},
  author={Cao, Yewen and Shao, Yulin},
  journal={arXiv preprint arXiv:2512.14424},
  year={2025}
}

@ARTICLE{11173628,
  author={Yin, Haoran and Tang, Yanqun and Ni, Yuanhan and Wang, Zulin and Chen, Gaojie and Xiong, Jun and Yang, Kai and Kountouris, Marios and Liang Guan, Yong and Zeng, Yong},
  journal={IEEE J. Sele. Areas Commun.}, 
  title={Ambiguity Function Analysis of {AFDM} Signals for Integrated Sensing and Communications}, 
  year={2026},
  volume={44},
  number={},
  pages={196-211},
  doi={10.1109/JSAC.2025.3611936}}

@article{SFDMtech,
  title={Stepped Frequency Division Multiplexing: A Jump-Free Continuous-Time {AFDM} Waveform},
  author={Cao, Yewen and Shao, Yulin},
  journal={arXiv:2605.11657},
  year={2026}
}

@article{shao2021federated,
  title={Federated edge learning with misaligned over-the-air computation},
  author={Shao, Yulin and G{\"u}nd{\"u}z, Deniz and Liew, Soung Chang},
  journal={IEEE Transactions on Wireless Communications},
  volume={21},
  number={6},
  pages={3951--3964},
  year={2021}
}

@article{shao2026embodied,
  title={Embodied Communication: Sensing-Induced Reliability Fields and Capacity Bounds},
  author={Shao, Yulin},
  journal={arXiv:2605.08284},
  year={2026}
}
